\documentclass{aa}
\usepackage[varg]{txfonts}
\usepackage{graphics}
\usepackage{graphicx}
\usepackage{color}
\usepackage{epsfig}
\usepackage{array}
\usepackage{ctable}
\usepackage{amssymb}
\usepackage{amsmath}
\usepackage{url}
\usepackage{lscape}
\usepackage{amsfonts}
\usepackage{mathpazo}
\usepackage{color}
\usepackage{fmtcount}
\usepackage{lipsum}
\usepackage{multirow}
\usepackage{subfig}
\usepackage{natbib}
\usepackage{threeparttable}
\bibpunct{(}{)}{;}{a}{}{,}

\begin{document}

\title{Solar differential rotation in the period 1964 - 2016 determined by the Kanzelh{\"o}he data set}

\newcommand{\angstrom}[1]{\AA}

\author{I. Poljan\v{c}i\'{c} Beljan\inst{1}
 \and R. Jurdana-\v{S}epi\'{c}\inst{1}
  \and R. Braj\v{s}a\inst{2}
   \and D. Sudar\inst{2}
    \and D. Ru\v{z}djak\inst{2}
     \and D. Hr\v{z}ina\inst{3}
      \and W. P{\"o}tzi\inst{4}
       \and A. Hanslmeier\inst{5}
        \and A. Veronig\inst{4,5}
         \and I. Skoki\'{c}\inst{6}
          \and H. W{\"o}hl\inst{7}}

\authorrunning{I. Poljan\v{c}i\'{c} Beljan et al.}
\titlerunning{Solar differential rotation determined by the Kanzelh{\"o}he data set}

\institute{Physics Department, University of Rijeka, Radmile Matej\v{c}i\'{c} 2, 51000 Rijeka, Croatia
\and Hvar Observatory, Faculty of Geodesy, University of Zagreb, Ka\v{c}i\'{c}eva 26, 10000 Zagreb, Croatia
\and Zagreb Astronomical Observatory, Opati\v{c}ka 22, 10000 Zagreb, Croatia
\and Kanzelh{\"o}he Observatory for Solar and Environmental Research, University of Graz, Kanzelh{\"o}he 19, 9521 Treffen am Ossiacher See, Austria
\and Institute for Geophysics, Astrophysics and Meteorology, Institute of Physics, University of Graz, Universit{\"a}tsplatz 5, 8010 Graz, Austria
\and Astronomical Institute of the Czech Academy of Sciences, Fri\v{c}ova 298, 25165 Ondřejov, Czech Republic
\and Kiepenheuer-Institut f{\"u}r Sonnenphysik, Sch{\"o}neckstr. 6, 79104 Freiburg, Germany}

\date{ Received /Accepted }

\abstract{Kanzelh{\"o}he Observatory for Solar and Environmental Research (KSO) provides daily multispectral synoptic observations of the Sun using several telescopes. In this work we made use of sunspot drawings and full disk white light CCD images.}
%However, it seems to be an underused data set, particularly regarding the use of sunspot drawings.
{The main aim of this work is to determine the solar differential rotation by tracing sunspot groups during the period 1964 - 2016, using the KSO sunspot drawings and white light images. We also compare the differential rotation parameters derived in this paper from the KSO with those collected fromf other data sets and present an investigation of the north - south rotational asymmetry.}
{Two procedures for the determination of the heliographic positions were applied: an interactive procedure on the KSO sunspot drawings (1964 - 2008, solar cycles nos. 20 - 23) and an automatic procedure on the KSO white light images (2009 - 2016, solar cycle no. 24). For the determination of the synodic angular rotation velocities two different methods have been used: a daily shift (DS) method and a robust linear least-squares fit (rLSQ) method. Afterwards, the rotation velocities had to be converted from synodic to sidereal, which were then used in the least-squares fitting for the solar differential rotation law. A comparison of the interactive and automatic procedures was performed for the year 2014.}
{The interactive procedure of position determination is fairly accurate but time consuming. In the case of the much faster automatic procedure for position determination, we found the rLSQ method for calculating rotational velocities  to be more reliable than the DS method. For the test data from 2014, the rLSQ method gives a relative standard error for the differential rotation parameter $B$ that is three times smaller than the corresponding relative standard error derived for the DS method. The best fit solar differential rotation profile for the whole time period is $\omega(b)$ = (14.47 $\pm$ 0.01) - (2.66   $\pm$ 0.10) $\sin^2b$ (deg/day) for the DS method and $\omega(b)$ = (14.50 $\pm$ 0.01) - (2.87 $\pm$ 0.12) $\sin^2b$ (deg/day) for the rLSQ method. A barely noticeable north - south asymmetry is observed for the whole time period 1964 - 2016 in the present paper. Rotation profiles, using different data sets, presented by other authors for 
the same time periods and the same tracer types, are in good agreement with our results.}
{The KSO data set used in this paper is in good agreement with the Debrecen Photoheliographic Data and Greenwich Photoheliographic Results and is suitable for the investigation of the long-term variabilities in the solar rotation profile. Also, the quality of the KSO sunspot drawings has gradually increased during the last 50 years.}


\maketitle

%**************************************************************************************************************************
%******************************** I N T R O D U C T I O N  *******************************************************************

\section{Introduction} \label{Introduction}

Solar rotation can be generally determined with three methods: the tracer
method, the spectroscopic method, and the helioseismology method \citep{2004suin.book.....S}. Sunspots and sunspot groups represent one of the most commonly used tracers in the literature. The main reason is the availability of long-term data sets from various observatories: the Greenwich Photoheliographic Results (GPR) data set \citep{balthwoehl1980,arevaloetal1982,balthvawo1986}; the Mt.Wilson data set \citep{howard1984,gilmanhow1984,hathawaywi1990}; the Extended Greenwich Results (EGR) data set \citep{pulktuo1998,javaraiah2003,zucczapp2003,javaraiahetal2005,javaraiahul2006,brajsaetal2006,brajsaetal2007,2014MNRAS.439.2377S}; the Kodaikanal data set \citep{guptasiho1999}; the Debrecen Photoheliographic Data (DPD)  \citep{Sudar2017}; and the Kanzelh{\"o}he Observatory for Solar and Environmental Research (KSO) data set \citep{1982AA...106..151L,lustig1983,Balth_Fangme1988,2014CEAB...38...87P,2016SoPh..291.3103P}. 

The KSO provides daily multispectral synoptic observations of the Sun using several telescopes \citep{2016ASPC..504..247V}. However, especially the sunspot drawings and white light images seem to be neglected by the scientific community. Previous studies performed on solar drawings from the KSO, in which sunpots and sunspot groups were analysed, involve data obtained through the year 1985 \citep{1982AA...106..151L,lustig1983,1984A&A...141..105L,1986A&A...160..277B,1986A&A...154..227H,1987A&A...172..332L,Balth_Fangme1988,1991A&A...249..528L}. There are a few papers that use more recent sunspot drawings \citep{2006A&A...447..735T,2010SunGe...5...52P,2011CEAB...35...59P,2014CEAB...38...87P}.  \citet{2006A&A...447..735T} provide the catalog of the hemispheric sunspot numbers and investigate the north - south activity asymmetries, but do not deal with the solar rotation.

\begin{figure}
   \centering
   \resizebox{12cm}{!}{\includegraphics[bb = 8 455 725 809]{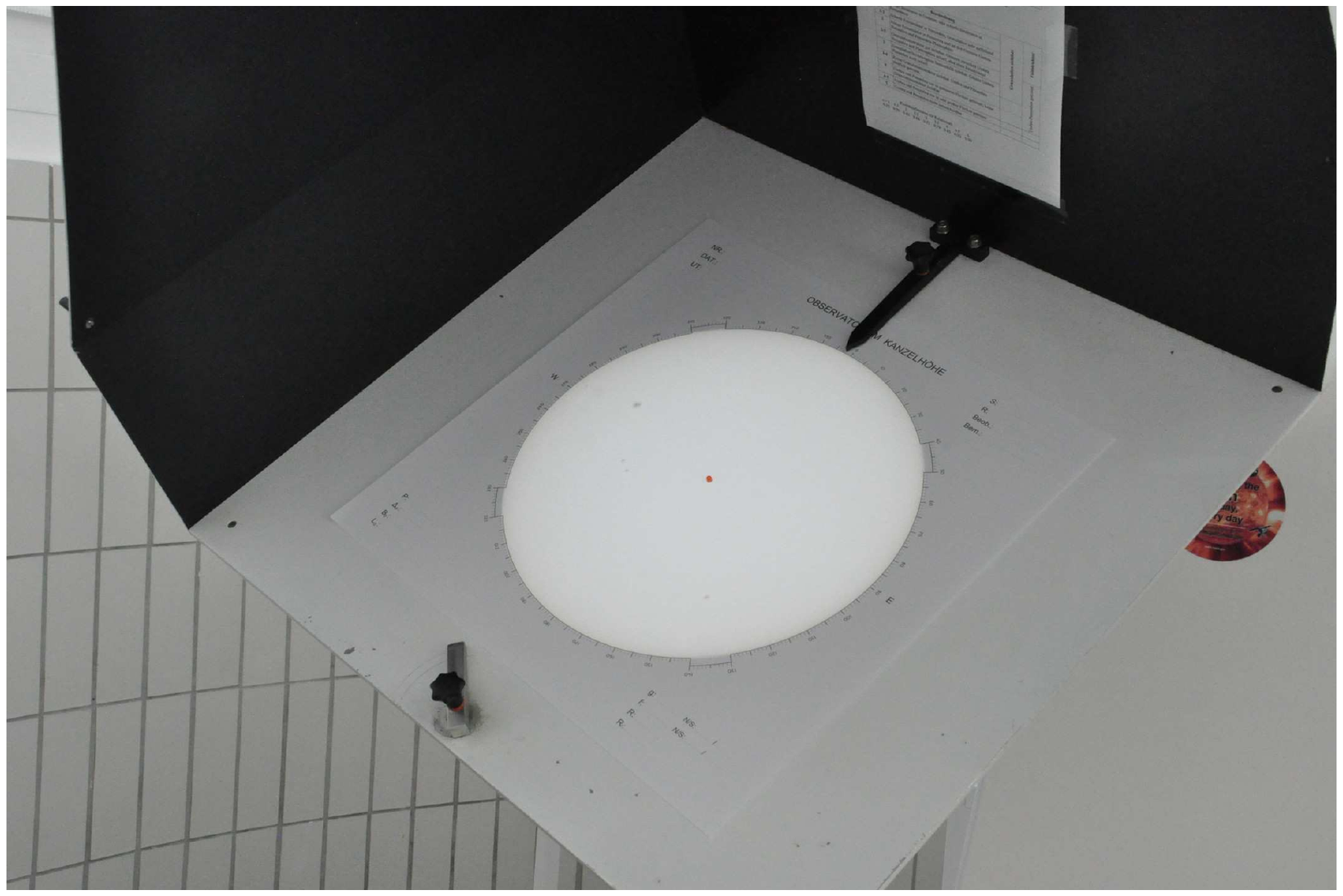}}\     \caption{Projection device for making the draft of sunspot drawings. The mirrors in the light path of the refractor have reversed the drawing, i.e., west is left. The enlargement of the solar disk to a diameter of about 25 cm corresponds to a magnification factor of about 100.}
     \label{Fig1}
\end{figure}

\citet{2010SunGe...5...52P}  checked the precision of the Solar Optical Observing Network/United States Air Force/National
Oceanic and Atmospheric Administration (SOON/USAF/NOAA) data set by comparing it with the GPR data set. The KSO data set (heliographic coordinates determined from the sunspot drawings for the years 1972 and 1993) were used as a reference for the comparison of sunspot position measurements. The SOON/USAF/NOAA data were found to be somewhat less accurate than the GPR data. \citet{2011CEAB...35...59P} expanded that analysis on several more data sets, again by using the KSO data set as a reference, for the comparison of sunspot position measurements. In almost all cases the latitude, longitude, and synodic angular velocity differences between the KSO and other data sets are the smallest for DPD, GPR, and SOON/USAF/NOAA. Since the GPR and SOON/USAF/NOAA data sets have already been widely used for the analysis of differential rotation of sunspot groups and its temporal variation, we decided that the next step will be to process  the DPD and KSO data sets. An analysis concerning the KSO data set is presented in this work, while an analysis of the DPD data set is made in \citet{Sudar2017}. \citet{2014CEAB...38...87P} calculated differential rotation parameters and showed differential rotation profiles for  solar cycles nos. 20 and 22 for the KSO data in a preliminary form.

Our analysis is a continuation of the investigation done by \citet{lustig1983}. We processed the KSO sunspot drawings for   solar cycles nos. 20 - 23 (1964 - 2008), and the KSO white light images for  solar cycle no. 24 (2009 - 2016). For the first time, the whole solar cycle no. 21 is examined using the KSO data (previously, it was just partially processed), as well as  solar cycles nos. 22-24.  Solar cycle no. 20 is where    our work and the work of \citet{lustig1983} overlap and is therefore suitable for comparison of the results. 

In this paper we study the solar differential rotation and we  make a comparison of the differential rotation parameters collected from different sources. In addition, we discuss the north-south asymmetry of the solar rotation profile.

%**********************************************************************************************************************
%**************************************************Instrumentation and measurements****************************************

\section{Instrumentation and measurements}

\begin{figure}
   \centering
   \resizebox{12cm}{!}{\includegraphics[bb = 8 255 775 819]{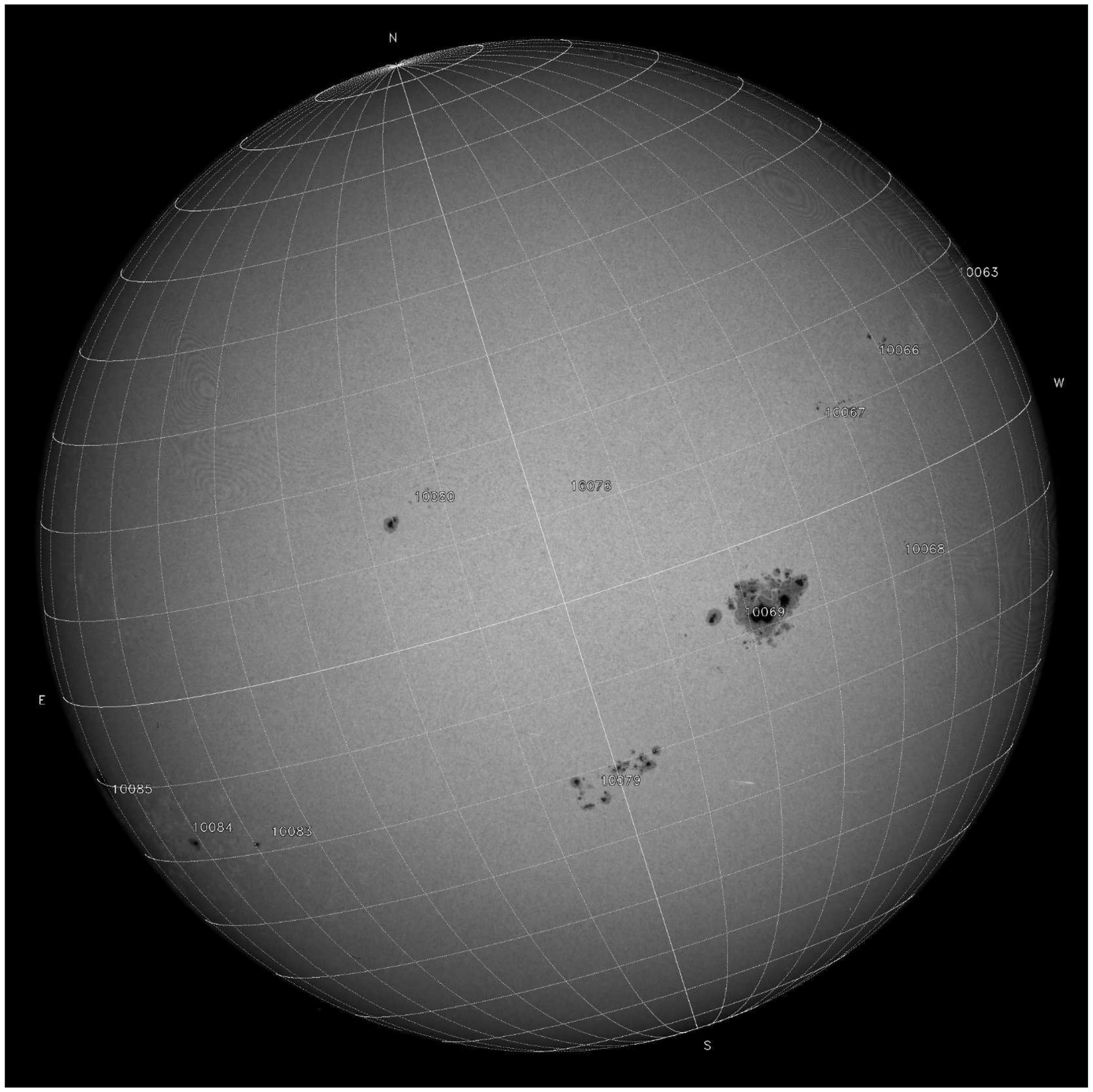}}\     \caption{Scanned white light image with added NOAA numbers and a grid that shows $P$ and $B_0$ angles. The image is not rotated, but an interface for de-rotating the images into the east-west direction is available (2002-08-19 06:10:41 UT).}
     \label{Fig2}
\end{figure}

The KSO belongs to the University of Graz, Austria.\footnotemark
\footnotetext{https://www.kso.ac.at/index\_en.php}
Sunspot drawings are made with a refractor ($d/f$ = 110/1650 mm) using the projection system that enlarges the disk image to 25 cm in diameter (see Fig.~\ref{Fig1}). The drawings are updated online almost every day, depending on weather conditions. In general, weather conditions are good and continuous operation yields about 300 days of observations per year. The KSO archive consists of more than 60 years of sunspot drawings \citep{2016SoPh..291.3103P}. \citet{2006SunGe...1b..21O} described the solar monitoring program at Kanzelh{\"o}he Observatory and the sunspot drawings in detail.

The white light images are made by a refractor ($d/f$ = 130/1950 mm) using a filter for 546 nm with a band pass of 10 nm. During the period 1989 - 2007 white light images were recorded on photographic film. Usually, three white light images per day were produced. Until recently, these images were accessible only on transparent sheet films, but   the scanning process is proceeding and they are now available online for the years after 1992 (see Fig.~\ref{Fig2}). The films are scanned with a photo scanner for transparent film material and as an output FITS and JPEG files are produced \citep{2010CEAB...34....1P}.

In July 2007 the system was replaced by the KSO Photosphere Digital Camera (KPDC), Pulnix TM-4100CL \citep{2010CEAB...34....1P}. To improve the image quality a 12 bit 2048 x 2048 pixel Pulnix RM-4200GE camera was installed in August 2015.

%**********************************************************************************************************************
%**************************************************Determination of the heliographic positions**********************************

\section{Determination of the heliographic positions}

Our investigation covers the time period 1964 - April 2016. We used two procedures to  determine  the heliographic positions of sunspot groups: an interactive procedure, using KSO sunspot drawings for the time period 1964 - 2008, and an automatic procedure, using KSO white light images for the time period 2009 - 2016.

In order to avoid solar limb effects leading to high position uncertainties, we limited the data to $\pm$58 deg in central meridian distance (CMD) which covers about 85\% of the projected solar radius \citep{balthvawo1986}. With this cutoff we obtained a sample of 12152 sunspot groups that correspond to some 70000 individual positions of sunspot groups. Recurrent sunspot groups are counted as many times as they appear.

\subsection{Interactive procedure (1964 - 2008, solar cycles nos. 20 - 23)}

A software package called Sungrabber\footnote{http://www.zvjezdarnica.hr/sungrabber/sungrabb.html}
\citep{2007CEAB...31..273H} that determines the position of tracers  is used for the interactive procedure. The area weighted 
centers of sunspot groups are estimated by the naked eye, thus giving more importance (by moving the center of gravity) to the umbra and penumbra that are more pronounced, i.e., occupy a larger area.

In order to find out if different observers could affect the measurements, we performed a test: two observers independently determined the positions of 10 single H- and J-type sunspot groups \citep[see][Fig.~3]{2010SunGe...5...52P}. Differences between measurements of the two observers are negligible (on average $\approx$ 0.1 deg, always less then 0.2 deg) when compared with the differences between various observatories ($\approx$ 0.5 deg). Based on these results all interactive coordinate determinations in the present work were done by two different observers using the Sungrabber software.

Sunspot groups were identified with the help of the GPR and DPD databases \citep{2016SoPh..291.3081B,2017MNRAS.465.1259G}. The software Sungrabber offers the possibility of loading the GPR or DPD sunspot group heliographic positions, as well as Greenwich or NOAA/USAF sunspot group numbers, and mark them on the sunspot drawing used.

\subsection{Automatic procedure (2009 - 2016,  solar cycle no. 24)}

Based on the algorithm by \citet{2011IAUS..273...51W} sunspot groups and their properties (size, umbra, penumbra, position) are identified by morphological image processing of KSO white light images.  This data is prepared every observing day by KSO and available over KSO ftp server\footnote{http://cesar.kso.ac.at/main/ftp.php} as raw fits images\footnote{ftp://ftp.kso.ac.at/phokaD/FITS/synoptic/} and data files\footnote{ftp://ftp.kso.ac.at/sunspots/drawings/automatic/}. Based on the assumption that images of better quality show more detail, we selected for each single day the fits file with the corresponding data file that contained the largest amount of information.

The data files provide the sunspot group center of gravity in pixel coordinates. Only the umbra pixels are used to calculate the center of gravity. For each sunspot group, a few separate umbra/penumbra pixel coordinates are also available, determined as the mean position of the umbra/penumbra pixels, which are only area weighted. 

From the fits header $P$ and $B_0$ were extracted, and from the corresponding data file solar radii and $X$ and $Y$ pixel coordinates were extracted. Using the calculations from \citet{meeus1991}, we calculated heliographic coordinates of sunspot groups. 

We also tried to apply the automatic method on white light images recorded on photographic film material (1989 - 2007), but the results were not very reliable. The calculated equatorial rotation velocities  ranged from less than 14 deg/day  to more than 15 deg/day. The reason for these uncertainties is the strong variation of the density (blackening) due to the developing procedure of the film material. These variations lead to some very faint images where the solar limb is very difficult to detect, leading to incorrect solar radii and as a consequence to incorrect sunspot positions.

%**********************************************************************************************************************
%**************************************************Determination of the rotation velocities**********************************

\section{Determining the rotation velocities}

In order to calculate the synodic angular rotation velocities we used two methods: {the daily-shift method} (DS), where  the synodic rotation velocities were calculated from the daily differences of the $CMD$ and the elapsed time $t$
\begin{equation}
\omega_{\mathrm{syn}} = \frac{\Delta CMD}{\Delta t} 
,\end{equation}and {the robust linear least-squares fit method} (rLSQ), where the synodic rotation velocities were calculated by fitting a line to the measured positions in time $CMD(t)$ for each tracer. The synodic rotation velocity corresponds to the slope of the fit. Since the measured data occasionally have outliers, due to false identification and other reasons, we used a robust fit, namely iteratively reweighted least-squares with Huber's t weighting function \citep{huber1981}. When applying the rLSQ method, the rotation velocities were calculated by fitting a line to at least three data points. If the number of position measurements, i.e., data points, is calculated for each sunspot group, the median value is 5, while the maximum value reaches 11. We note that even in the cases where only three data points is used by the rLSQ method, it is still  more than the two data points used when applying the DS method.

%///////////////////////////////  TABLE 1 - Aut vs. int comparison, results ///////////////////////
\begin{table}[!ht]
\begin{center}
\caption{Results of fittings of the KSO data for the year 2014 to rotational velocity profile Eq. 2. The differential rotation parameters are calculated by the interactive (int) and the automatic (aut) procedures and both hemispheres together. All calculations are done for DS and rLSQ methods separately and passed through $\pm$58 deg $CMD$ filter and 8 - 19 deg/day velocity filter. The sidereal parameters and their standard errors are expressed in deg/day. $N_{vl}$ is the number of  calculated rotation velocities.}\label{Tab1}
\begin{tabular}[c]{>{\centering}m{0.50cm}>{\centering}cccc}\hline\hline\noalign{\smallskip}

Row&Method&$A$ &  $B$ &$N_{vl}$ \\\hline\noalign{\smallskip}

1&int, DS       &               14.55 $\pm$     0.06    &       -2.14   $\pm$         0.83&989        \\
2&aut, DS       &               14.42   $\pm$   0.08    &       -0.98   $\pm$         1.11&831        \\\hline\noalign{\smallskip}
3&int, rLSQ     &               14.62   $\pm$   0.08    &       -2.52   $\pm$         1.11&218        \\
4&aut, rLSQ     &               14.64   $\pm$   0.09    &       -3.40   $\pm$         1.32&180        \\
\hline\noalign{\smallskip}

\end{tabular}

\end{center}
\end{table}

%/////////////////////////////// FIGURE 3 ///////////////////////

\begin{figure}[!ht]
  \centering
    \resizebox{7cm}{!}{\includegraphics[bb = 140 51 720 553]{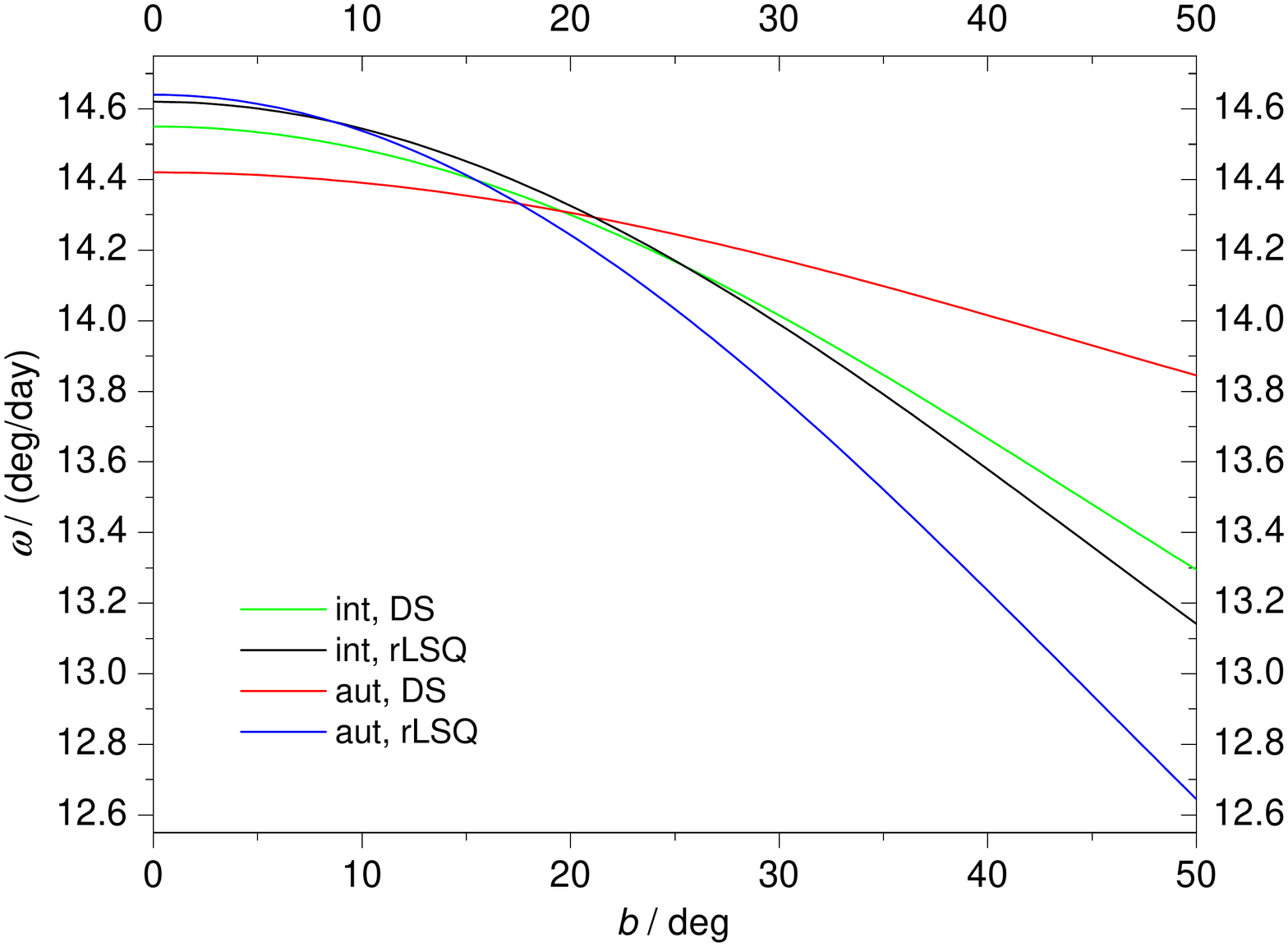}}\
    \caption{ Differential rotation profiles for the year 2014 (see Table~\ref{Tab1}). Sidereal rotation velocity is denoted by $\omega$ and heliographic latitude by $b$.}
    \label{Fig3}
\end{figure}

Angular rotation velocities were converted from synodic to sidereal  using the procedure described by \citet{1995SoPh..159..393R} and by \citet{2002SoPh..206..229B}, which was improved by \citet{2014SoPh..289.1471S}. The calculated sidereal velocities, $\omega$, were used in the least-squares fitting to the solar differential rotation law
\begin{equation}
\omega(b) = A+B\sin^2b
,\end{equation}
where $b$ is the heliographic latitude, and $A$ and $B$ are the solar differential rotation parameters.
In both methods, DS and rLSQ, we assigned the velocity to the latitude and time of the first measurement of position. When average latitudes are used, the false meridional flows can be detected \citep{2005AstL...31..706O}, but for rotation analysis it probably represents a negligible effect.

%**********************************************************************************************************************
%**************************************************RESULTS****************************************

\section{Results}
\subsection{Comparison of automatic and interactive procedures}

%/////////////////////////////// FIGURE 4 ///////////////////////
\begin{figure}[!ht]
   \centering
   \resizebox{8cm}{!}{\includegraphics[bb = 88 35 755 563]{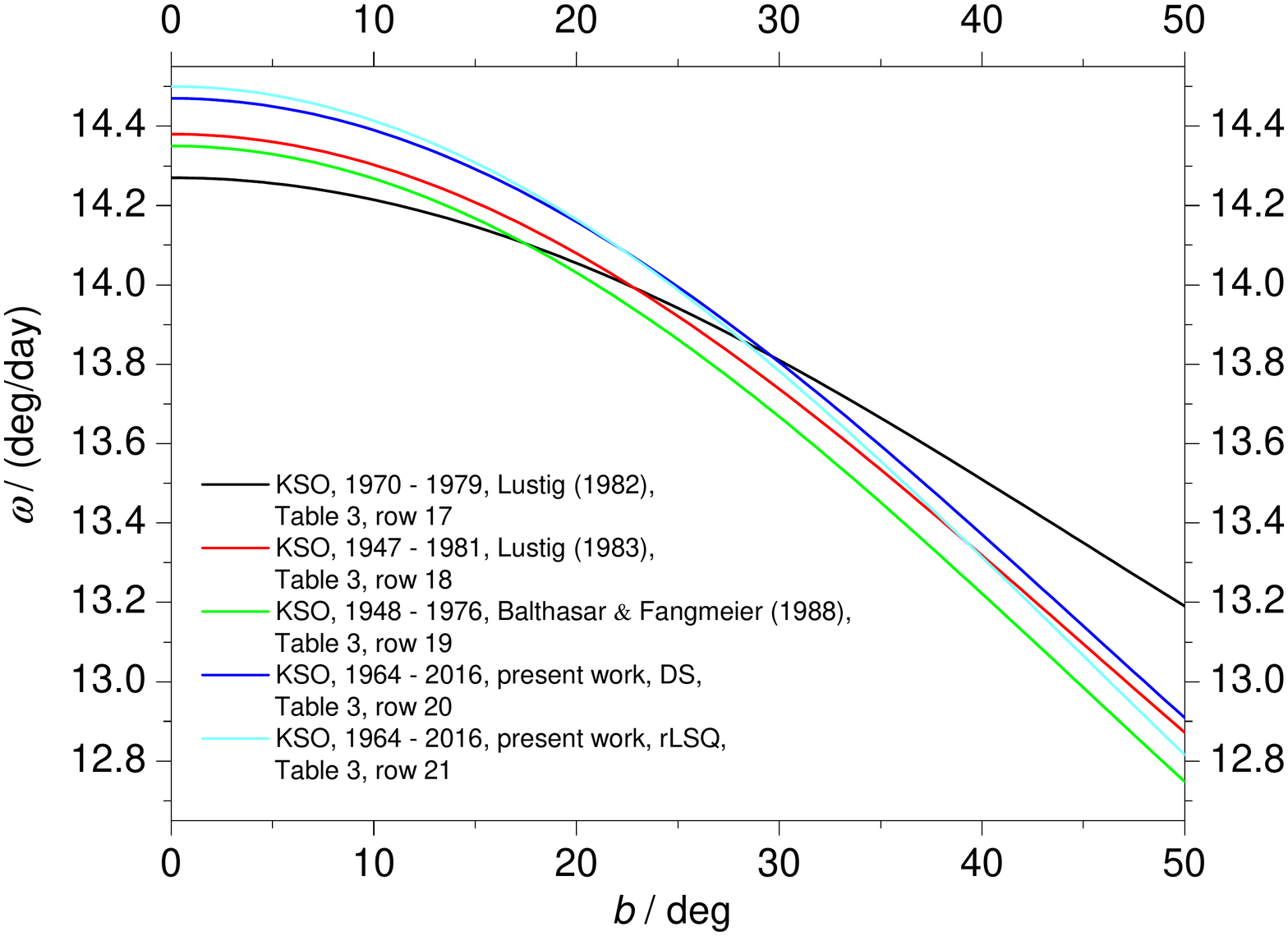}}\
     \caption{ Differential rotation profiles calculated by different authors for the KSO data set and both hemispheres together (see corresponding rows in Table~\ref{Tab3}). Sidereal rotation velocity is denoted by $\omega$ and heliographic latitude by $b$.}
     \label{Fig4}
\end{figure}

A direct comparison of the interactive and the automatic procedure can only be presented for the year 2014, the only year for which both procedures were applied. This year was selected as it represents the maximum of  solar cycle no. 24 after a deep minimum, when almost no sunspots and thus data points were available, and the first maximum covered by the KSO Photosphere Digital Camera. We note that the years near the solar minimum usually lead to large statistical errors as the data set becomes very small, e.g., the year 2009 with 260 spotless days.

The differential rotation parameters $A$ and $B$ determined for the year 2014 with
both procedures and the velocities obtained with DS and rLSQ methods for both hemispheres together are presented in Table 1. Corresponding differential rotation 
profiles are shown in Fig.~\ref{Fig3}. More details are given in Sect. \ref{Automatic vs. interactive procedure, DS vs. rLSQ method}.

\subsection{Differential rotation from the KSO data}

%///////////////////////////////  TABLE 2 - Results of fittings ///////////////////////

\begin{table*}[!ht]
\begin{center}
\caption{Results of fittings of the KSO data to rotational velocity profile Eq. 2. Differential rotation parameters are calculated separately for solar cycles nos. 20 - 24, for the part of the data set using the interactive procedure (covering solar cycles nos. 20 - 23) - int (all), and for the whole data set using the interactive and automatic procedure together (covering solar cycles nos. 20 - 24) -  int+aut (all). All results are calculated separately via the DS and rLSQ methods covering the whole Sun (i.e., both hemispheres together) - N+S, the northern hemisphere -  N,  and the southern hemisphere - S. All calculations are exposed to $\pm$58 deg $CMD$ filter and 8 - 19 deg/day velocity filter. The sidereal parameters and their standard errors are expressed in deg/day. $N_{vl}$ is the number of  calculated rotation velocities.}\label{Tab2}
\begin{tabular}[c]{>{\centering}cccccccc}\hline\hline\noalign{\smallskip}
Row&Method&Cycle &Time period&Hemisphere&$A$ &  $B$ & $N_{vl}$\\\hline\noalign{\smallskip}

1&int, DS       &       20& 1964.8 - 1976.3     &N+S&   14.45   $\pm$   0.02    &       -3.05   $\pm$   0.27    &       5319\\
2&int, DS       &       21 &1976.3 - 1986.7     &N+S&   14.50   $\pm$   0.02    &       -2.47   $\pm$   0.20    &       7486\\
3&int, DS       &       22 &1986.7 - 1996.4     &N+S&   14.44   $\pm$   0.02    &       -2.81   $\pm$   0.17    &       8400\\
4&int, DS       &       23 &1996.4  - 2008.9    &N+S&   14.50   $\pm$   0.02    &       -2.54   $\pm$   0.17    &       8204\\
5&aut, DS       &       24 &2008.9  - 2016.3    &N+S&   14.48   $\pm$   0.04    &       -2.61   $\pm$   0.45    &       3701\\
6&int (all), DS &       20 - 23 &1964.8 - 2008.9        &N+S&   14.47   $\pm$         0.01    &       -2.66   $\pm$   0.10    &       29409\\
7&int + aut (all), DS   &        20 - 24 &1964.8 - 2016.3       &N+S&   14.47   $\pm$         0.01    &       -2.66   $\pm$   0.10    &       33110
\\\hline\noalign{\smallskip}
8&int, DS       &       20 &1964.8 - 1976.3     &N&     14.45   $\pm$   0.03    &       -3.17   $\pm$   0.35    &       2845    \\
9&int, DS       &       21 &1976.3 - 1986.7     &N&     14.49   $\pm$   0.03    &       -2.26   $\pm$   0.29    &       3675    \\
10&int, DS      &       22 &1986.7 - 1996.4     &N&     14.43   $\pm$   0.03    &       -3.13   $\pm$   0.25    &       3882    \\
11&int, DS      &       23 &1996.4  - 2008.9    &N&     14.50   $\pm$   0.03    &       -2.79   $\pm$   0.25    &       3838    \\
12&aut, DS      &       24 &2008.9  - 2016.3    &N&     14.47   $\pm$   0.06    &       -2.48   $\pm$   0.68    &       1893    \\
13&int (all), DS        &       20 - 23 &1964.8 - 2008.9        &N&     14.47   $\pm$   0.02    &       -2.83   $\pm$   0.14    &       14240   \\
14&int + aut (all), DS  &       20 - 24 &1964.8 - 2016.3        &N&     14.47   $\pm$   0.02    &       -2.82   $\pm$   0.14    &       16133
\\\hline\noalign{\smallskip}
15&int, DS      &       20 &1964.8 - 1976.3     &S&     14.44   $\pm$   0.03    &       -2.78   $\pm$   0.45    &       2474            \\
16&int, DS      &       21 &1976.3 - 1986.7     &S&     14.52   $\pm$   0.03    &       -2.67   $\pm$   0.28    &       3811            \\
17&int, DS      &       22 &1986.7 - 1996.4     &S&     14.44   $\pm$   0.03    &       -2.52   $\pm$   0.24    &       4518            \\
18&int, DS      &       23 &1996.4  - 2008.9    &S&     14.49   $\pm$   0.03    &       -2.33   $\pm$   0.24    &       4366            \\
19&aut, DS      &       24 &2008.9  - 2016.3    &S&     14.49   $\pm$   0.06    &       -2.74   $\pm$   0.62    &       1808            \\
20&int (all), DS        &       20 - 23 &1964.8 - 2008.9        &S&     14.47   $\pm$   0.02    &       -2.50   $\pm$   0.14    &       15169           \\
21&int + aut (all), DS  &       20 - 24 &1964.8 - 2016.3        &S&     14.47   $\pm$   0.01    &       -2.51   $\pm$   0.13    &       16977           
\\\hline\hline\noalign{\smallskip}

22&int, rLSQ    &       20 &1964.8 - 1976.3     &N+S&   14.45   $\pm$   0.03    &       -3.06   $\pm$   0.33    &       1216\\
23&int, rLSQ    &       21 &1976.3 - 1986.7     &N+S&   14.53   $\pm$   0.03    &       -2.66   $\pm$   0.26    &       1717\\
24&int, rLSQ    &       22 &1986.7 - 1996.4     &N+S&   14.50   $\pm$   0.02    &       -3.17   $\pm$   0.20    &       1763\\
25&int, rLSQ    &       23 &1996.4  - 2008.9    &N+S&   14.51   $\pm$   0.03    &       -2.79   $\pm$   0.22    &       1757\\
26&aut, rLSQ    &       24 &2008.9  - 2016.3    &N+S&   14.52   $\pm$   0.05    &       -2.60   $\pm$   0.52    &       787\\
27&int (all), rLSQ      &       20 - 23 &1964.8 - 2008.9        &N+S&   14.50   $\pm$   0.01    &       -2.89   $\pm$   0.12    &       6453\\
28&int + aut (all), rLSQ        &        20 - 24 &1964.8 - 2016.3       &N+S&   14.50   $\pm$   0.01    &       -2.87   $\pm$   0.12    &       7240
\\\hline\noalign{\smallskip}
29&int, rLSQ    &       20 &1964.8 - 1976.3     &N&     14.46   $\pm$   0.04    &       -3.26   $\pm$   0.39    &       640\\
30&int, rLSQ    &       21 &1976.3 - 1986.7     &N&     14.51   $\pm$   0.04    &       -2.35   $\pm$   0.38    &       836\\
31&int, rLSQ    &       22 &1986.7 - 1996.4     &N&     14.50   $\pm$   0.03    &       -3.50   $\pm$   0.27    &       803\\
32&int, rLSQ    &       23 &1996.4  - 2008.9    &N&     14.52   $\pm$   0.04    &       -2.89   $\pm$   0.34    &       813\\
33&aut, rLSQ    &       24 &2008.9  - 2016.3    &N&     14.41   $\pm$   0.07    &       -1.87   $\pm$   0.87    &       405\\
34&int (all), rLSQ      &       20 - 23 &1964.8 - 2008.9        &N&     14.50   $\pm$   0.02    &       -3.01   $\pm$   0.17    &       3092\\
35&int + aut (all), rLSQ        &       20 - 24 &1964.8 - 2016.3        &N&     14.49   $\pm$   0.02    &       -2.94   $\pm$   0.17    &       3497
\\\hline\noalign{\smallskip}
36&int, rLSQ    &       20 &1964.8 - 1976.3     &S&     14.43   $\pm$   0.05    &       -2.76   $\pm$   0.57    &       576\\
37&int, rLSQ    &       21 &1976.3 - 1986.7     &S&     14.54   $\pm$   0.04    &       -2.94   $\pm$   0.36    &       881\\
38&int, rLSQ    &       22 &1986.7 - 1996.4     &S&     14.49   $\pm$   0.03    &       -2.83   $\pm$   0.29    &       960\\
39&int, rLSQ    &       23 &1996.4  - 2008.9    &S&     14.50   $\pm$   0.03    &       -2.71   $\pm$   0.30    &       944\\
40&aut, rLSQ    &       24 &2008.9  - 2016.3    &S&     14.66   $\pm$   0.07    &       -3.59   $\pm$   0.63    &       382\\
41&int (all), rLSQ      &       20 - 23 &1964.8 - 2008.9        &S&     14.49   $\pm$   0.02    &-2.77 $\pm$   0.17    &       3361\\
42&int + aut (all), rLSQ        &       20 - 24 &1964.8 - 2016.3        &S&     14.51   $\pm$   0.02    &-2.81 $\pm$   0.17    &       3743
\\\hline\hline\noalign{\smallskip}

\end{tabular}

\end{center}
\end{table*}

Using both procedures, interactive and automatic, five solar cycles (20 - 24) from 1964 until 2016 were used to obtain the differential rotation parameters. Both solar hemispheres were treated together and separately.

With only a $CMD$ cutoff we obtained a sample of 41125 (33817 via DS and 7308 via rLSQ) calculated sidereal rotation velocities. Rotational velocity outliers resulting from misidentification of sunspot groups in subsequent images can be filtered out by applying the velocity filter 8-19 deg/day \citep{2002A&A...392..329B,2003A&A...404.1117V,2014MNRAS.439.2377S,2015A&A...575A..63S}. After selecting only sidereal rotation velocities higher than 8 deg/day and lower than 19 deg/day, the number of calculated sidereal rotation velocities was reduced to 40480 (33213 via DS and 7267 via rLSQ). This means that 1.82\% of the calculated velocities were eliminated using the DS method and 0.56\% using the rLSQ method. This also means that 51\% of the reduced calculated sidereal rotation velocities belong to the northern hemisphere, while 49\% belong to the southern hemisphere (valid for both DS and rLSQ). 

In Table~\ref{Tab2} we present the results of fittings for  solar cycles nos. 20 - 24 separately (DS: rows 1 - 5 for N+S, rows 8 - 12 for N, rows 15 - 19 for S; rLSQ: rows 22 - 26 for N+S, rows 29 - 33 for N, rows 36 - 40 for S); the results for the part of the data set using the interactive procedure covering solar cycles nos. 20 - 23 (DS: rows 6, 13, 20; rLSQ: rows 27, 34, 41) and the results for the whole data set using the interactive and automatic method together covering solar cycles nos. 20 - 24 (DS: rows 7, 14, 21; rLSQ: rows 28, 35, 42). All results are calculated using  both methods  (DS and rLSQ), for both hemispheres together (N+S), and for the northern hemisphere (N) and the southern hemisphere (S) separately. The years specified in Table~\ref{Tab2} (Time period) represent the starting and ending epochs of the corresponding solar cycles. They are taken from \citet{2009A&A...496..855B}, except the beginning of  solar cycle no. 24, which was taken from  \citet{sidc}, Royal Observatory of Belgium, Brussels\footnote{http://www.sidc.be/silso/news004}.

\subsection{Comparison with other data}

%/////////////////////////////// TABLE 3 - references ///////////////////////

\begin{table*}[!p]
\begin{center}
\caption{Values of differential rotation parameters $A$ and $B$ and their standard errors (in deg/day) collected from different sources using only sunspots and sunspot groups as tracers. N denotes the northern hemisphere, S denotes the southern hemisphere, and N+S both hemispheres together. In Col. 2: GPR - Greenwich Photoheliographic Results, EGR - Extended Greenwich Results, KSO - Kanzelh{\"o}he Observatory for Solar and Environmental Research data, CS - Carrington \& Sp{\"o}rer data, DPD - Debrecen Photoheliographic Results.}\label{Tab3}
\begin{tabular}
%[c]{>{\centering}m{0.5cm}>{\centering}m{1.8cm}c>{\centering}m{1.7cm}>{\centering}m{1.7cm}ccc}\hline\hline\noalign{\smallskip}
[c]{>{\centering}m{0.30cm}>{\centering}m{2.4cm}c>{\centering}m{1.7cm}>{\centering}m{2.4cm}ccc}\hline\hline\noalign{\smallskip}

Row&Data set    & Time   &Hemisphere& $A$  & $B$&References     \\\hline\noalign{\smallskip}

1&GPR\tablefootmark{a}      &1874 - 1976 &N+S   &14.551 $\pm$ 0.006     &-2.87  $\pm$ 0.06    &1      \\
2&GPR\tablefootmark{a}  &1874 - 1902&N+S        &14.63  $\pm$ 0.01      &-2.70 $\pm$ 0.16      &2\\
3&      GPR\tablefootmark{a}    &       1879 - 1975     &       N+S     &       14.522  $\pm$   0.005   &       -2.66   $\pm$   0.04    &3\\
4&      GPR\tablefootmark{b}    &       1880 - 1976     &       N+S     &       14.37   $\pm$   0.01    &       -2.59   $\pm$   0.16&4  \\
5&      GPR\tablefootmark{a}    &       1883 - 1893     &       N+S     &       14.63   $\pm$ \textcolor{white}{0.0}- &       -2.69   $\pm$   \textcolor{white}{0.0}-  &5\\
6&GPR\tablefootmark{a}  &1940 - 1968&N+S        &14.53  $\pm$ 0.01      &-2.83 $\pm$ 0.08&6    \\
7&      GPR\tablefootmark{a}    &       1948 - 1976     &       N+S     &       14.52   $\pm$ \textcolor{white}{0.0}- &       -2.84   $\pm$ \textcolor{white}{0.0}-   &5\\
8&EGR\tablefootmark{a}  &1878 - 2011&N+S        &14.49  $\pm$ 0.01      &-2.64 $\pm$ 0.05&7    \\
9&      EGR\tablefootmark{a}    &       1976 - 2002     &       N+S     &       14.457  $\pm$   0.009   &       -2.17   $\pm$   0.07&3  \\
10&     EGR\tablefootmark{c}    &       1874 - 1996     &       N+S     &       14.531  $\pm$   0.003   &       -2.747  $\pm$   0.048   &8\\
11&     CS\tablefootmark{c}     &       1853 - 1893     &       N+S     &       14.475  $\pm$   0.011   &       -2.710  $\pm$   0.165   &8\\
12&     Sp{\"o}rer\tablefootmark{a}     &       1883 - 1893     &       N+S         &       14.50   $\pm$   \textcolor{white}{0.0}- &       -2.41   $\pm$   \textcolor{white}{0.0}- &5\\
13&     Abastumani\tablefootmark{d}     &       1950 - 1990     &       N+S         &       14.73   $\pm$   0.06    &       -2.07   $\pm$   0.51&9  \\
14&Mt. Wilson\tablefootmark{d}  &1921 - 1982&N+S        &14.522 $\pm$ 0.004     &-2.84 $\pm$ 0.04      &10\\
15&Mt. Wilson\tablefootmark{a}  &1921 - 1982&N+S        &14.393 $\pm$ 0.010     &-2.95 $\pm$ 0.09&10   \\
16&DPD\tablefootmark{a} &1974 - 2016&N+S        &14.50  $\pm$0.01       &-2.54 $\pm$ 0.07&11   \\
17&     KSO\tablefootmark{c}    &       1970 - 1979     &       N+S     &       14.27   $\pm$   0.02    &       -1.84   $\pm$   0.12&12 \\
18&     KSO\tablefootmark{c}    &       1947 - 1981     &       N+S     &       14.38   $\pm$   0.01    &       -2.57   $\pm$   0.07&13 \\
19&     KSO\tablefootmark{a}    &       1948 - 1976     &       N+S     &       14.35   $\pm$ \textcolor{white}{0.0}- &       -2.73 $\pm$     \textcolor{white}{0.0}- &5\\
20& KSO\tablefootmark{a}& 1964 - 2016& N+S&     14.47   $\pm$   0.01    &       -2.66   $\pm$   0.10&15 \\
21& KSO\tablefootmark{a}& 1964 - 2016& N+S&     14.50   $\pm$   0.01    &       -2.87   $\pm$   0.12&16 \\

\hline\noalign{\smallskip}

22&GPR\tablefootmark{a}       &1874 - 1976 & N&14.54    $\pm$ 0.01      &-2.88  $\pm$ 0.08    &1      \\
23&GPR\tablefootmark{a} &1874 - 1902&N  &14.65  $\pm$ 0.02      &-2.89 $\pm$ 0.16&2  \\
24&     GPR\tablefootmark{a}    &       1879 - 1975     &       N       &       14.531  $\pm$   0.005   &       -2.63   $\pm$   0.06&3  \\
25&GPR\tablefootmark{a} &1940 - 1968&N  &14.51  $\pm$ 0.01      &-2.69 $\pm$ 0.11&6  \\
26&     EGR\tablefootmark{a}    &       1976 - 2002     &       N       &       14.462  $\pm$   0.015   &       -2.13   $\pm$   0.10&3  \\
27&     EGR\tablefootmark{c}    &       1874 - 1996     &       N       &       14.53   $\pm$   0.01    &       -2.69   $\pm$   0.07&8  \\
28&     CS\tablefootmark{c}     &       1853 - 1893     &       N       &       14.48   $\pm$   0.02    &       -3.02   $\pm$   0.26&8  \\
29&     KSO\tablefootmark{c}    &       1947 - 1981     &       N       &       14.38   $\pm$   0.01    &       -2.70   $\pm$   0.09&13 \\
30&     KSO\tablefootmark{c}    &       1964 - 1976     &       N       &       14.40   $\pm$   0.02    &       -2.75   $\pm$   0.24&14 \\
31& KSO\tablefootmark{a}& 1964 - 2016& N&       14.47   $\pm$   0.02    &       -2.82   $\pm$   0.14    &17\\
32& KSO\tablefootmark{a}& 1964 - 2016& N&       14.49   $\pm$   0.02    &       -2.94   $\pm$   0.17&18 \\
\hline\noalign{\smallskip}

33&GPR\tablefootmark{a}       &1874 - 1976 & S&14.56    $\pm$ 0.01      &-2.85  $\pm$ 0.09    &1      \\
34&GPR\tablefootmark{a} &1874 - 1902&S  &14.61  $\pm$ 0.02      &-2.56 $\pm$ 0.16&2  \\
35&     GPR\tablefootmark{a}    &       1879 - 1975     &       S       &       14.517  $\pm$   0.005   &       -2.68   $\pm$   0.05&3  \\
36&GPR\tablefootmark{a} &1940 - 1968&S  &14.55  $\pm$ 0.01      &-3.00 $\pm$ 0.13&6  \\
37&     EGR\tablefootmark{a}    &       1976 - 2002     &       S       &       14.448  $\pm$   0.015   &       -2.20   $\pm$   0.10&3\\ 
38&     EGR\tablefootmark{c}    &       1874 - 1996     &       S       &       14.54   $\pm$   0.01    &       -2.81   $\pm$   0.07&8\\ 
39&     CS\tablefootmark{c}     &       1853 - 1893     &       S       &       14.48   $\pm$   0.02    &       -2.51   $\pm$   0.22&8\\ 
40&     KSO\tablefootmark{c}    &       1947 - 1981     &       S       &       14.38   $\pm$   0.01    &       -2.34   $\pm$   0.11&13 \\
41&     KSO\tablefootmark{c}    &       1964 - 1976     &       S       &       14.37   $\pm$   0.02    &       -2.48   $\pm$   0.27&14 \\
42& KSO\tablefootmark{a}& 1964 - 2016& S&       14.47   $\pm$   0.01    &       -2.51   $\pm$   0.13&19 \\
43& KSO\tablefootmark{a}& 1964 - 2016& S&       14.51   $\pm$   0.02    &       - 2.81    $\pm$   0.17&20 \\
\hline\hline\noalign{\smallskip}

\end{tabular}
\end{center}

\tablefoot{
Type of  tracers:\\
\tablefoottext{a}{sunspot groups}
\tablefoottext{b}{stable recurrent sunspot groups}
\tablefoottext{c}{sunspots and sunspot groups}
\tablefoottext{d}{sunspots}
}

\tablebib{
(1)~\citet{balthvawo1986}; (2) \citet{arevaloetal1982}; (3) \citet{javaraiah2003}; (4) \citet{2002SoPh..206..229B}; 
(5) \citet{Balth_Fangme1988}; (6) \citet{balthwoehl1980}; (7) \citet{2014MNRAS.439.2377S}; (8) \citet{pulktuo1998}; (9) \citet{2002SoPh..206..219K}; (10) \citet{1984ApJ...283..373H}; (11) \citet{Sudar2017}; (12) \citet{1982AA...106..151L}; (13) \citet{lustig1983}; (14) \citet{lustig1983}, cycle 20; (15) present work, Table~\ref{Tab2}, row 7, DS; (16) present work, Table~\ref{Tab2}, row 28, rLSQ; (17) present work, Table~\ref{Tab2}, row 14, DS; (18) present work, Table~\ref{Tab2}, row 35, rLSQ; (19) present work, Table~\ref{Tab2}, row 21, DS; (20) present work, Table~\ref{Tab2}, row 42, rLSQ.
}
\end{table*}

%/////////////////////////////// FIGURE 5 ///////////////////////
\begin{figure*}[!ht]
   \centering
   \resizebox{8cm}{!}{\includegraphics[bb = 280 45 585 553]{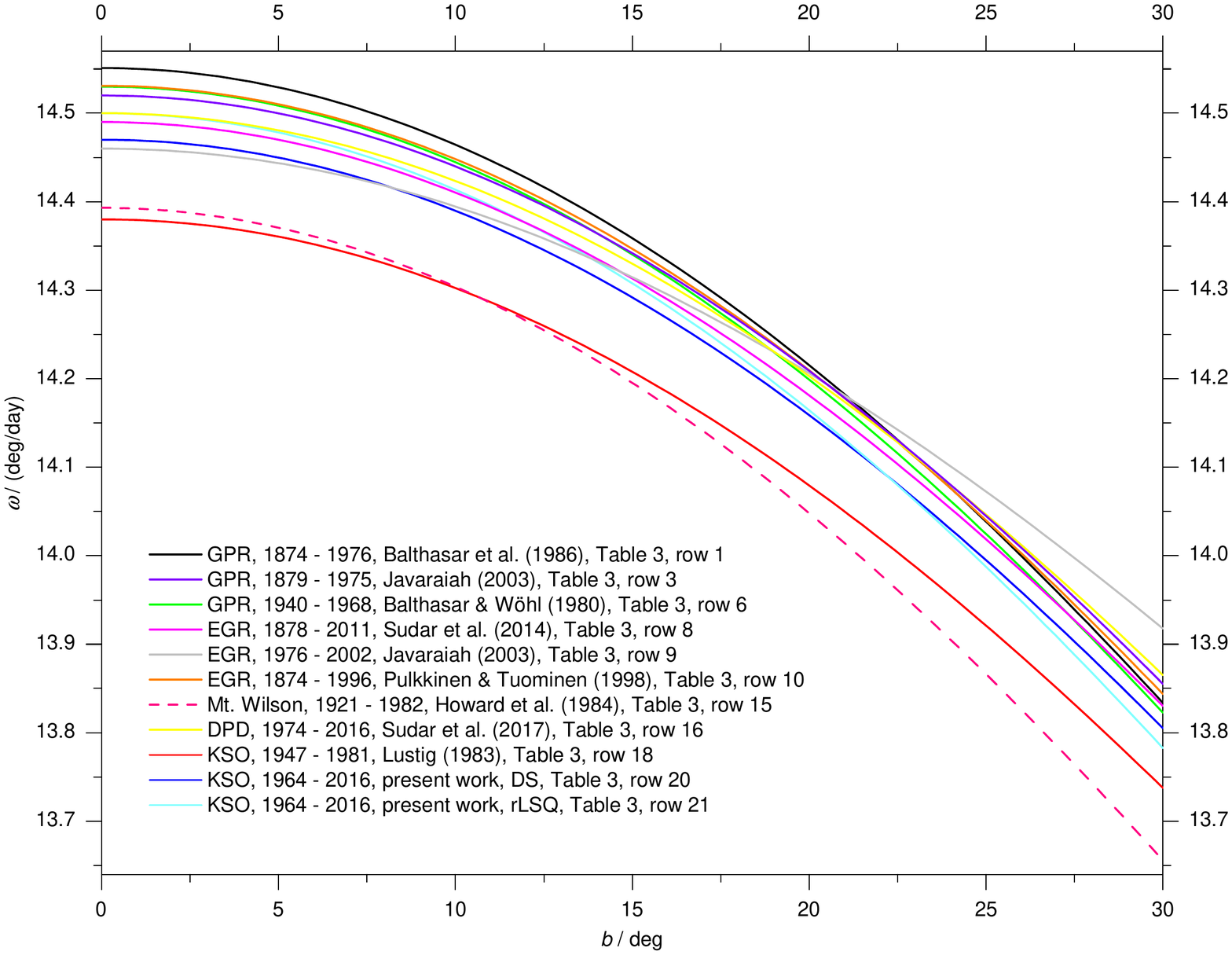}}\
     \caption{ Differential rotation profiles calculated by different authors for several data sets (GPR, EGR, Mt. Wilson, DPD, KSO) and both hemispheres together (see corresponding rows in Table~\ref{Tab3}). Sidereal rotation velocity is denoted by $\omega$ and heliographic latitude by $b$.}
     \label{Fig5}
\end{figure*}

In this section we present a comparison of the differential rotation parameters $A$ and $B$ collected from different sources, mostly those already mentioned in Sect. \ref{Introduction} as interesting (GPR, EGR, DPD, KSO) dealing only with sunspots and sunspot groups as tracers (Table~\ref{Tab3}). Table~\ref{Tab3} consists of three parts: the upper part (rows 1 - 21) gives the results for both solar hemispheres together, the middle part (rows 22 - 32) gives the results for the northern hemisphere, and the lower part (rows 33 - 43) gives the results for the southern hemisphere. Figure~\ref{Fig4} only shows the differential rotation profiles  for the results derived from the KSO data set (Table~\ref{Tab3}, rows 17 - 21). Figure~\ref{Fig5} shows the differential rotation profiles for several data sets and both hemispheres together (see corresponding rows in Table~\ref{Tab3}), using only sunspot groups (Note (a)) or sunspots and sunspot groups (Note (c)) as tracers. The differential rotation profiles derived from the KSO data set \citep{lustig1983} (Fig.~\ref{Fig5}, red line) and the Mt. Wilson data set \citep{1984ApJ...283..373H} (Fig.~\ref{Fig5}, pink dashed line) show significantly lower values when compared to other results. More details are given in Section \ref{A comparison with other data sets}.

\subsection{North - south asymmetry of the solar rotation}

The differential rotation parameters derived in the present paper for the whole time period (1964 - 2016) and for the northern and southern hemispheres separately are given in Table~\ref{Tab2} (DS method - rows 14 and 21; rLSQ method - rows 35 and 42). Corresponding differential rotation profiles are shown in Fig.~\ref{Fig6}. More details are given in Section \ref{The north - south asymmetry of the solar rotation}.

%****************************************************************************************************************************
%***********************************DISCUSSION ******************************************

\section{Discussion}
\subsection{Automatic vs. interactive procedure, DS vs. rLSQ method} \label{Automatic vs. interactive procedure, DS vs. rLSQ method}

If we compare the results for the year 2014 determined by the interactive and automatic procedures and by the DS method (Table~\ref{Tab1}, rows 1 and 2), for both differential rotation parameters the results are within the 1 common $\sigma$, which also holds for the rLSQ results (Table~\ref{Tab1}, rows 3 and 4). However, both procedures, the interactive and the automatic, show large standard errors, especially for parameter $B$. A comparison of the standard errors in Table~\ref{Tab1} (rows 1 and 2, rows 3 and 4) between the two procedures reveals that the automatic procedure is less accurate but can process more data. The application of the interactive procedure is rather time consuming and the results presented here are a combination of the efforts of two co-authors over almost two years of measurements. 

It is obvious that the differential rotation profiles for the rLSQ method coincide better (Fig.~\ref{Fig3}, black and blue lines) because the equatorial rotation velocities are almost the same and the gradients of the solar rotation do not differ as in the DS case. In the case of the automatic procedure, the rLSQ method for calculating rotational velocities (Fig.~\ref{Fig3}, blue line) is much more reliable than the DS method (Fig.~\ref{Fig3}, red line), which can be clearly seen by a comparison of the corresponding differential rotation parameters  in Table~\ref{Tab1} (rows 2 and 4). For example,   parameter $B$ calculated via  the DS method yields $B = (-0.98 \pm 1.11)$ deg/day, while for the rLSQ method $B = (-3.40 \pm 1.32)$ deg/day. If relative standard errors are compared, the rLSQ method yields a value that is three times smaller than the corresponding value for the DS method.  

However, if we look at the results obtained by applying only the interactive procedure (Table~\ref{Tab2}, rows 6, 13, 20, 27, 34, 41) and compare them with the results of the interactive and automatic procedures together covering the whole time period (Table~\ref{Tab2}, rows 7, 14, 21, 28, 35, 42), we see that the inclusion of the results calculated by the automatic procedure, for  solar cycle no. 24, does not affect the ``whole time period'' result. In other words, the int (all) and int + aut (all) results from the corresponding rows in Table~\ref{Tab2}, which are listed above, are almost identical in all cases for both the DS and rLSQ methods and for both hemispheres. Thus, when the rotation velocities calculated by the automatic procedure are combined with the rotation velocities calculated by the interactive procedure for longer periods of time, the effect of the lower accuracy of the automatic procedure has no statistically significant influence on the final best fit differential rotation profile. Hence, we combine both procedures if we analyse the whole period from 1964 to 2016. No separation between an interactive procedure before 2009 and an automatic procedure after 2009 has to be made.
 
For white light images  recorded on photographic film material before 2007, which are scanned, the automatic procedure did not yield reliable results, and we were forced to process this data using the interactive procedure and sunspot drawings. So, for the years before 2007, it is possible to apply the automatic procedure after elimination of the possible errors that appear during the scanning procedure. However, for the years after 2007, the automatic procedure represents a useful tool for processing the white light images.

If we look at Table~\ref{Tab2} and compare the DS and rLSQ
measurements derived for the whole periods (both hemispheres - rows 7 and 28; northern hemisphere - rows 14 and 35; southern hemisphere - rows 21 and 42), the errors for both differential rotation parameters in all cases (northern,  southern, or both hemispheres) are within the 1 common $\sigma$. In some cases there are very small (i.e., almost negligible) differences. This means that the results of the two different methods used to  determine the rotation velocities for the whole time period (daily shift - DS and robust linear least-squares fit - rLSQ) can be considered  almost identical.

\subsection{Comparison with other data sets} \label{A comparison with other data sets}

In \citet{2011CEAB...35...59P} a comparison of sunspot position measurements  for several data sets was made, separately for single H and J and complex sunspot groups types. The mean absolute differences between the KSO and DPD positions are lower in all cases -- except for the longitude difference for complex sunspot groups -- than the mean absolute difference between the KSO and GPR positions. As the KSO and GPR comparison was made for the year 1972, it may be possible that the KSO data is influenced by the erroneous solar image radii \citep{1984SoPh...91...55B}, thus causing a larger difference than that between the KSO and the DPD. The KSO and DPD comparison was made for the year 1993, and was no longer influenced by the previously mentioned unwanted effect, because the KSO observing and reduction procedures were improved after the 1980s \citep{1984SoPh...91...55B}.

We compared the differential rotation parameters derived in the present work from KSO for both the DS and the rLSQ methods (Table~\ref{Tab3}, rows 20 and 21) with values for the EGR \citep{2014MNRAS.439.2377S} (Table~\ref{Tab3}, row 8) and values for the DPD \citep{Sudar2017}(Table~\ref{Tab3}, row 16), which were also obtained by tracing sunspot groups and covering almost
the same time period. Values for the differential rotation parameter $A$  are within the 1 common $\sigma$, while values for the differential rotation parameter $B$   match within the 2 common $\sigma$. It is well known that the GPR represents a  homogeneous data set with  high accuracy \citep{balthwoehl1980,arevaloetal1982,2013SoPh..288..117W,2013SoPh..288..141W,2013SoPh..288..157E}, as well as the DPD as its continuation \citep{Sudar2017}. Therefore, we feel free to conclude that the KSO data used in this paper (from 1964 till present) is of  comparable accuracy to the DPD or GPR data sets, and suitable for the investigation of the solar rotation, although there is some indication of lower accuracy in the past, especially before the 1960s \citep{1984SoPh...91...55B}.    

Except for the already mentioned overlapping with \citet{2014MNRAS.439.2377S,Sudar2017} rotation results (Table~\ref{Tab3}, rows 8 and 16), in most of the  other studies listed in Table~\ref{Tab3} the differential rotation parameters are in agreement with our results (Table~\ref{Tab3}, rows 20 and 21); the exceptions are  the GPR result for reccurent sunspot groups \citep{2002SoPh..206..229B} (Table~\ref{Tab3}, row 4); the Mt. Wilson result concerning only sunspot groups as tracers \citep{1984ApJ...283..373H} (Table~\ref{Tab3}, row 15); the Abastumani result \citep{2002SoPh..206..219K} (Table~\ref{Tab3}, row 13); and the KSO results \citep{1982AA...106..151L,lustig1983,Balth_Fangme1988} (Table~\ref{Tab3}, rows 17 - 19).

%/////////////////////////////// FIGURE 6 ///////////////////////
\begin{figure}[!ht]
   \centering
   \resizebox{8cm}{!}{\includegraphics[bb = 58 25 775 581]{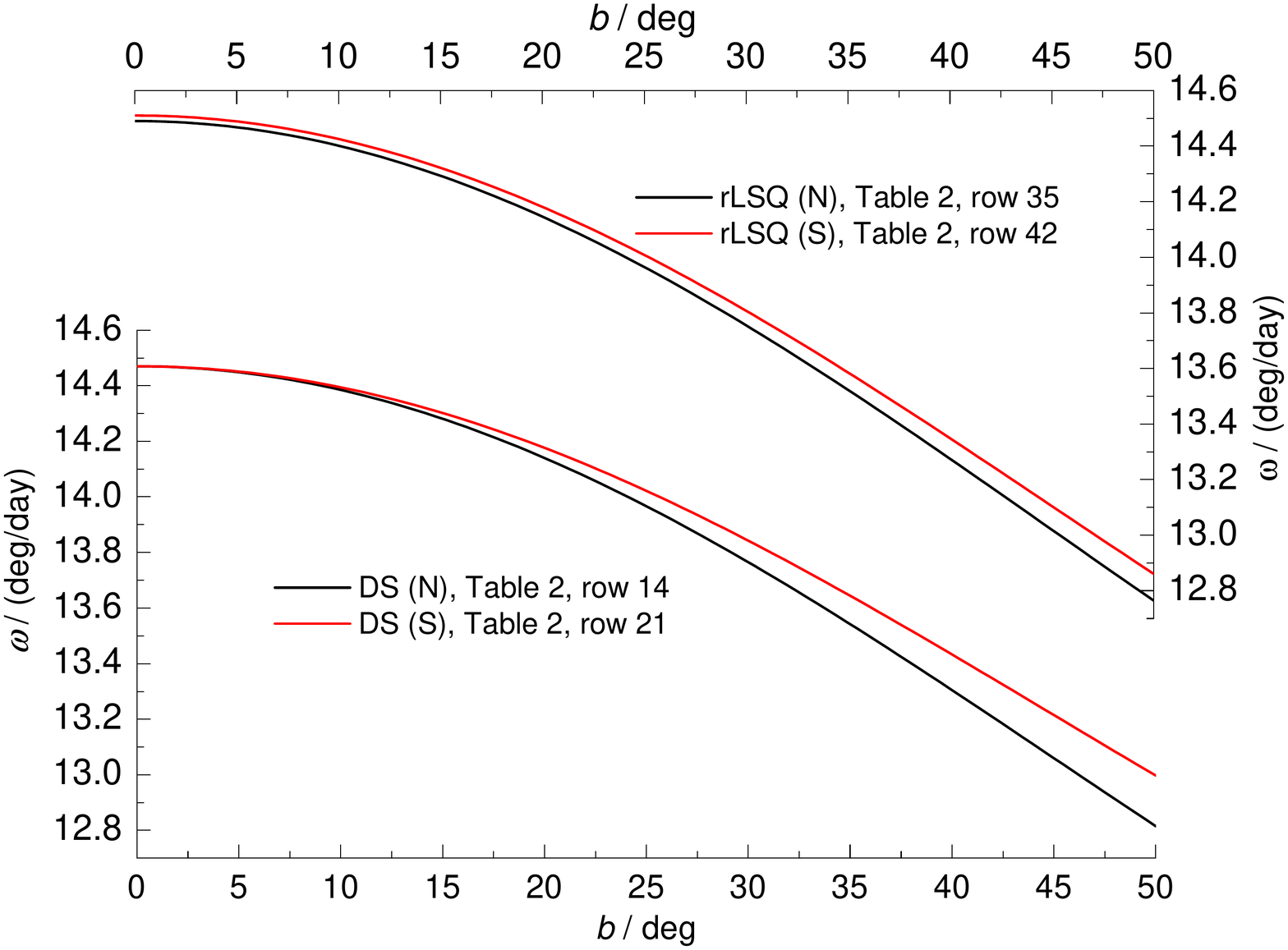}}\     \caption{ North - south asymmetry for the whole time period 1964 - 2016 and for DS and rLSQ methods (see also the corresponding rows in Table~\ref{Tab2}). Sidereal rotation velocity is denoted by $\omega$ and heliographic latitude by $b$. N - northern hemisphere, S - southern hemisphere.}
     \label{Fig6}
\end{figure}

According to \citep{2004SoPh..221..225R,2005SoPh..229...35R} recurrent sunspots and recurrent sunspot groups show a slower rotation, which may be the reason for the low rotation values of the
GPR given by \citet{2002SoPh..206..229B} (Table~\ref{Tab3}, row 4). It is interesting that the lower value of the equatorial velocity $A$ was also obtained  for the Mt. Wilson data for sunspot groups \citep{1984ApJ...283..373H} (Table~\ref{Tab3}, row 15), but it is not clear what could be the cause. A higher value of the equatorial velocity, as well as the lower value of the differential rotation gradient was obtained with the Abastumani data \citep{2002SoPh..206..219K} (Table~\ref{Tab3}, row 13) compared to our analysis (Table~\ref{Tab3}, rows 20 and 21). This result, however, shows higher standard errors in comparison to other results in Table~\ref{Tab3}.

The equatorial rotation rates $A$ derived from the KSO data \citep{1982AA...106..151L,lustig1983,Balth_Fangme1988} (Table~\ref{Tab3}, rows 17 - 19) for the years before the 1980s are significantly lower than our analysis (Table~\ref{Tab3} - rows 20 and 21) by about 0.15 deg. A large difference between differential rotation parameters $B$ is also noticed. \citet{Balth_Fangme1988} analysed the GPR data for the same time period (Table~\ref{Tab3}, row 7) and also noted the disagreement with the KSO data, confirming that the KSO data before the 1980s are affected by some systematic errors, already examined in \citet{1984SoPh...91...55B}. 

\begin{figure}[!ht]
   \centering
   \resizebox{8cm}{!}{\includegraphics[bb = 65 40 765 579]{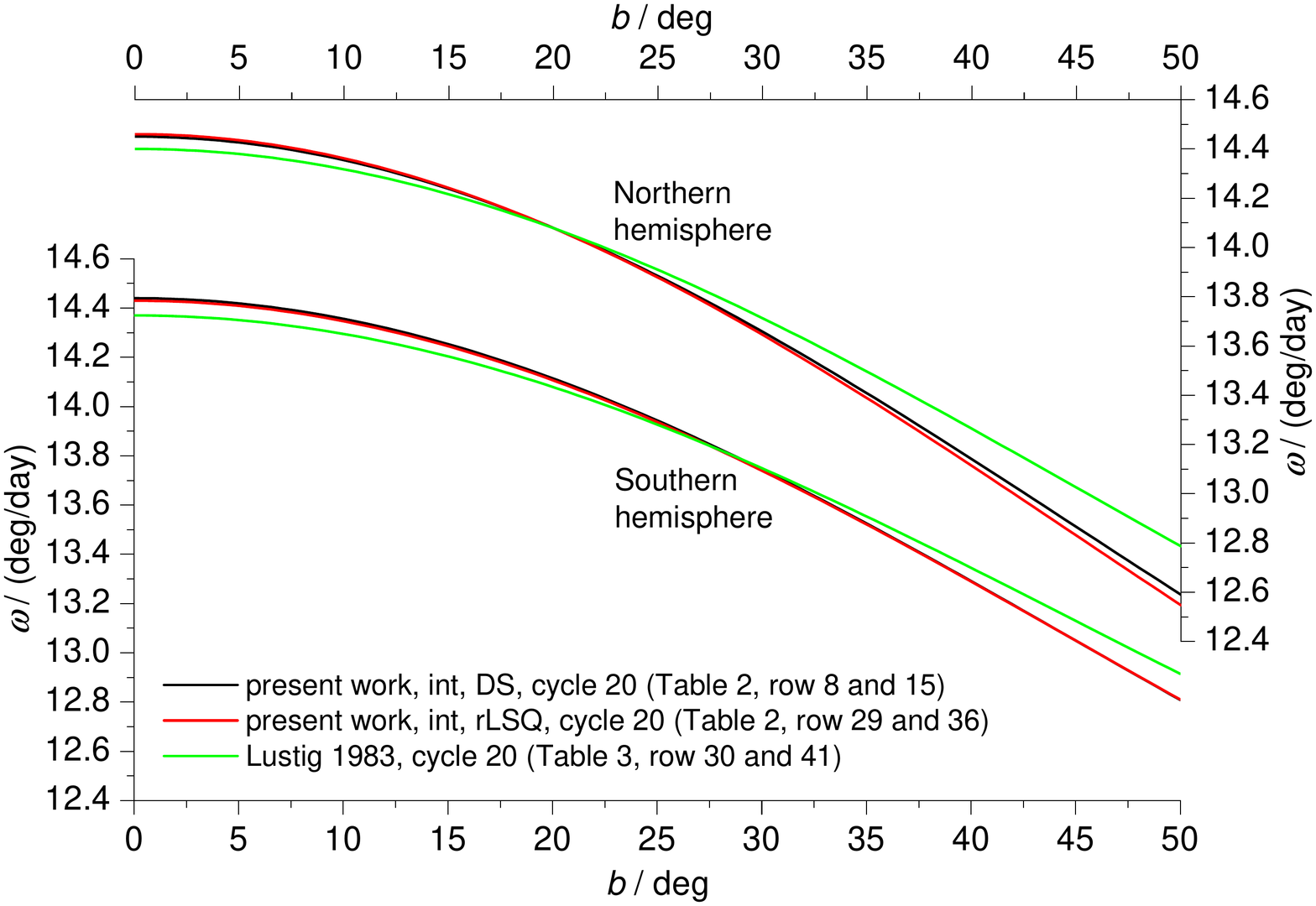}}\     \caption{ Comparison between the present paper and \citet{lustig1983} differential rotation profiles calculated for  solar cycle no. 20 (1964 - 1976) and for the northern and southern hemisphere separately, see corresponding rows in Tables~\ref{Tab2}~and~\ref{Tab3}. The sidereal rotation velocity is denoted by $\omega$ and heliographic latitude by $b$.}
     \label{Fig7}
\end{figure}

Since our investigations were conceived as a continuation of the investigations done in \citet{lustig1983}, it is important to compare them also for the overlapping period of time (1964 - 1976, solar cycle no. 20). \citet{lustig1983} used the Stonyhurst disk overlaying technique to determine the heliographic coordinates of sunspot groups by superimposing the disk on sunspot drawing. For  solar cycle no. 20, \citet{lustig1983} provides the solar rotation parameters only for the northern and southern hemispheres. They are listed in Table~\ref{Tab3} (rows 30 and 41). The corresponding solar rotation parameters derived in the present paper are listed in Table~\ref{Tab2} (rows 8 and 15 for DS, rows 29 and 36 for rLSQ). All these rotation profiles are shown in  Fig.~\ref{Fig7}. The results for both differential rotation parameters and both methods, the DS and the rLSQ, are within the 1 common $\sigma$. Only the differential rotation parameter $A$ for the DS method and southern hemisphere falls within the 2 common $\sigma$ with the corresponding result from \citet{lustig1983}. This comparison is  evidence that the reliability of the KSO data set has been improved, not only for the period after the  1980s, but  indeed from 1960 onwards.

\subsection{ North - south asymmetry of the solar rotation} \label{The north - south asymmetry of the solar rotation}

%/////////////////////////////// FIGURE 7 ///////////////////////

If we compare the DS measurements derived for the whole time period (1964 - 2016) and for the northern and the southern hemispheres separately (Table~\ref{Tab2}, rows 14 and 21), we see that the measurements for differential rotation parameter $A$ are within the 1 common $\sigma$, i.e.,  the difference is very small and statistically insignificant. Differential rotation parameter $B$ shows the difference between the two solar hemispheres, statistically significant on 2 common $\sigma$ level (Table~\ref{Tab2}, rows 14 and 21). Concerning the rLSQ measurements (Table~\ref{Tab2}, rows 35 and 42), we see that the measurements for both differential rotation parameters are within the 1 common $\sigma$. 

In \citet{lustig1983}, statistically insignificant differences were discovered for equatorial velocities (the parameter $A$) between the northern and southern hemispheres, while the difference in the gradients of the differential rotation (the parameter $B$) is 0.36 deg/day. It is interesting that in present work for the DS method this difference remains almost the same ($B_{N}-B_{S}=0.31$ deg/day). 

For both methods, the DS and the rLSQ, the southern solar hemisphere rotates a little bit faster, which can also be clearly seen in Fig.~\ref{Fig6}. Therefore, a barely noticeable north - south asymmetry is observed for the whole time period 1964 - 2016 in the present work.

%****************************************************************************************************************************
%***********************************CONCLUSIONS ******************************************

\section{Conclusions}

We have determined the solar differential rotation by tracing sunspot groups during the period 1964 - 2016. We have used two procedures to   determine the heliographic positions: an interactive one on the KSO sunspot drawings  (1964 - 2008, solar cycles nos. 20 - 23) and an automatic one on the KSO white light images  (2009 - 2016, solar cycle no. 24). For the first time, the whole solar cycle no. 21 was investigated with the KSO data, as well as solar cycles nos. 22 - 24. Applying a $CMD$ cutoff at $\pm$58 deg we obtained a sample of 12152 sunspot groups which correspond to approximately 70000 individual sunspot positions. The synodic angular rotation velocities were determined using two different methods, the DS and rLSQ, and then converted to the sidereal velocities. A sample of 41125 calculated sidereal rotation velocities was obtained. Calculated sidereal velocity values have been used in the least-squares fitting to the solar differential rotation law. 

The interactive and automatic procedures comparison was performed for the year 2014. It yielded 1 common $\sigma$ coincidence for both differential rotation parameters and both the DS and rLSQ methods. However, in both procedures, interactive and automatic, there are large standard errors for the differential rotation parameters, especially for the differential rotation parameter $B$. If the comparison were performed for the entire cycle, it would yield better results as the number of calculated velocities and latitudinal coverage is higher in that case. For the automatic procedure the standard errors are even larger, indicating lower accuracy of the automatic procedure in comparison to the interactive one. Also, in the case of automatic procedure, the rLSQ method for calculating rotational velocities is much more reliable than the DS method.  For the test data from 2014, the rLSQ method yields a relative standard error for the differential rotation parameter $B$ that is three times smaller than that of the DS method. But, when the rotation velocities calculated by the automatic procedure are combined with the rotation velocities calculated by the interactive procedure for longer periods of time, the effect of the lower accuracy of the automatic procedure has no statistically significant influence on the final best fit differential rotation profile. This allows us to discuss interactive procedure before 2009 and automatic procedure after 2009 together as if  only one procedure has been applied. 

The two different methods used for the determination of the rotation velocities (daily shift - DS and robust linear least-squares fit - rLSQ) were also compared for the whole time period, and showed 1 common $\sigma$ coincidence for both differential rotation parameters.

Inspection for the north - south asymmetry showed that for both methods, the DS and the rLSQ, the southern solar hemisphere rotates a little bit faster. Thereby, a barely noticeable north - south asymmetry is observed for the whole time period 1964 - 2016 in the present paper, very similar to previous observations for the time period 1947 - 1981 \citep{lustig1983}.

The comparison of calculated rotation profiles derived from the KSO data in this work and from all other studies collected from different sources yielded a conclusion about the KSO accuracy after the  1960s. The KSO data used in this paper for the time period from 1964 to the present is of a comparable accuracy with the DPD or GPR data sets, and suitable for the investigation of the solar rotation. We cannot guarantee that for the years prior to 1960 the results would be of the same accuracy because there are some indications about lower accuracy in the past \citep{1984SoPh...91...55B}.   

The quality of the KSO sunspot drawings has gradually increased over the last 50 years. In general, the interactive procedure of position determination is fairly accurate, but has the drawback of being very time consuming, so the much faster automatic procedure of position determination was developed. However, in the case of the automatic procedure, the rLSQ method for calculating rotational velocities is much more reliable than the DS method.

The main conclusion of this paper is that KSO provides a valuable data set with  satisfactory accuracy. Although there were some systematic errors  in the past, this data set is well suited to long-term studies and therefore still
continued.

We plan to continue our analysis of the KSO data. The next step is presenting the temporal variation of the differential rotation and the relationship between the solar rotation and activity, as well as an analysis of meridional motions and horizontal Reynolds stress.

\begin{acknowledgements}

This work was supported in part by the Croatian Science Foundation under the project 6212 \textquotedblleft Solar and Stellar Variability\textquotedblright\ and in part by the University of Rijeka under  project number 13.12.1.3.03. It
has also received funding from the SOLARNET project (312495, 2013-2017),
which is an Integrated Infrastructure Initiative (I3) supported by FP7 Capacities
Programme. IPB thanks the Kanzelh{\"o}he Solar Observatory for the hospitality during her stay at Kanzelh{\"o}he. 

\end{acknowledgements}

\bibliographystyle{aa.bst}
\bibliography{Ref_Kanz}

\end{document}